%% file: llncs.tex
\documentclass{llncs}
\usepackage{makeidx}  
\usepackage{url}
\usepackage{times}
\usepackage{latexsym}
\usepackage{amsmath}
\usepackage{graphicx}
\usepackage{tabu}
\usepackage{float}
\usepackage{booktabs}
\usepackage{algorithm}
\usepackage[noend]{algpseudocode}
\usepackage{multirow}
\usepackage{enumitem}
\usepackage{multicol}
\usepackage[linewidth=1pt]{mdframed}
\usepackage{booktabs}
\usepackage{graphicx}
\usepackage{url}
\usepackage{tikz}
\usepackage{lipsum}
\usepackage[symbol]{footmisc}
\usepackage{hyperref}
\hypersetup{
    colorlinks=true,
    linkcolor=blue,
    filecolor=magenta,      
    urlcolor=cyan,
}
\begin{document}
\newcommand*\samethanks[1][\value{footnote}]{\footnotemark[#1]}
%
\title{Inflo: News Categorization and Keyphrase Extraction for Implementation in an Aggregation System}
\titlerunning{Inflo}

\author{Pranav A\thanks{These two authors had joint contribution, the demo link and supplementary materials can be found at \href{https://www.dropbox.com/s/kw6wsnogrjyf2s9/ecir.pdf?dl=0}{this link}, emails are \texttt{cs.pranav.a@gmail.com, nsukiennik@ust.hk, panhui@cse.ust.hk}}, Nick Sukiennik\samethanks \and Pan Hui}
\institute{SyMLab, Hong Kong University of Science and Technology, Hong Kong}

\maketitle              

\begin{abstract}
The work herein describes a system for automatic news category and keyphrase labeling, presented in the context of our motivation to improve the speed at which a user can find relevant and interesting content within an aggregation platform. A set of 12 discrete categories were applied to over 500,000 news articles for training a neural network, to be used to facilitate the more in-depth task of extracting the most significant keyphrases. The latter was done using three methods: statistical, graphical and numerical, using the pre-identified category label to improve relevance of extracted phrases. The results are presented in a demo in which the articles are pre-populated via News API, and upon being selected, the category and keyphrase labels will be computed via the methods explained herein. 
\end{abstract}
\input{main.tex}

\bibliography{ref}
\bibliographystyle{splncs}
\end{document}

%% file: main.tex
\section{Introduction}
Accessing relevant news content on the web is a complex issue which has been addressed by aggregation media and a plethora of aggregation apps. 
Typical news aggregation apps allow users to select their topics of interest upon signing up for the app, like Flipboard \cite{flipboard} \cite{lumi.news}. Then they are presented with news that comes from a finite list of sources, according to their pre-selected topics of interest. This type of news navigation is not optimal in that does not take into account the fact that a user’s interests may be more specific than the topics offered. 

On the other hand, some aggregators learn about the users’ preferences over time and develop their recommendation algorithm to provide similar content, leading to an echo chamber effect \cite{singer_2011}. Overall, the current solutions lead to a phenomenon wherein the content presented is either A) too general, based solely on broad categories, or B) too specific, based on recommendation algorithms. 

For this demonstration, we have developed a news labeling system which sorts a news article into a set of predefined, discrete categories, as well as extracts its most significant keyphrases. In this demonstration, we have implemented our labeling results into a user-interface which pulls up the most recent news articles across some popular sources (via News API \cite{newsapi}), and when clicked, analyzes the text having been scraped via Python (Newspaper module \cite{newspaper3k}), displaying the automatic instantaneous results of our labeling algorithms. Such results lay the groundwork for what could be further elaborated into a content aggregation system wherein the output topics and keyphrases could be used as filtering tools, allowing for more efficient navigation of news within an aggregation platform.

\section{Preliminaries}
The pipeline of our demo comprises 1) finding an appropriate category for the news article, 2) extracting keyphrases with the help of category specific and entity based document frequencies, 3) finding similar articles based on list of extracted keywords. 
This section focuses on preparation of the dataset required by our algorithms to work.

\textbf{Dataset Collection:} 
With the help of various news based APIs, especially New York Times API \cite{nytimes}, CNN News API \cite{cnn} and News API \cite{newsapi}, we collected 500024 news articles from 2000 to 2018.
We decided to define 12 discrete news categories: Regional Politics, Sports, Entertainment, International Relations, Science, Business, War and Conflicts, Law and Order, Technology, US, World and Miscellaneous topics.
There were approx. 20000-50000 articles per category.
In order to better allow the models to understand vocabulary and styles occurring in recent articles, we doubled the dataset obtained from 2017-2018. 
We kept 80 \% articles for training, 10 \% for validation, and 10 \% for testing purposes.
Many articles did not have categories explicitly stated, and categories across news publications can vary; therefore, we had to apply one of the discrete categories defined above to each article manually. To do so, we allocated all articles of a given tag (from metadata) into the most relevant broad category (among those we have defined). 

\textbf{Document Frequency Computation:} 
For each news article, we have used two document frequency computations: 1) Phrase based and 2) Named-entity based. 
Both of these are category-specific.
That means, if our classifier predicts an article as "Sports", it would refer to our pre-computed sports frequency counts for phrases and entities.
The motivation behind this is to incorporate category-specific vocabulary for keyword extraction.
For entity-based document frequency computation, we used SpaCy \cite{spacy2}. First, we extracted entities for each article. Following that, we computed document frequencies on extracted entities to create a separate corpus of entities for each category. 
This enables the models to incorporate common entities which occur within that news category.

\section{News Classification}
Text classification has been one of the central problems in the NLP (Natural Language Processing) field.
Deep learning models, especially CNN (Convolutional Neural Networks) have been very effective for text classification for most general purposes \cite{kim2014convolutional} \cite{zhang2015character}.
However, these models often require huge datasets, computational power, and days to converge.
Transfer Learning for Deep Learning involves using a pre-trained model and fine-tuning for customized purposes \cite{bengio2012deep}.
At the time of writing, ULMFiT (Universal Language Model Fine-tuning for Text Classification) is the current state-of-the-art for text classification with transfer learning \cite{howard2018universal}.
The usage of this model avoids the requirement of a large dataset and computational power, resulting in higher accuracy within shorter amount of time.
Their model requires fine-tuning the language model and classifier on the target dataset.
A language model is a probabilistic distribution for a sequence of words, and AWD-LSTM (ASGD Weight-Dropped LSTM) has been used as a main architecture for the language model \cite{merity2017regularizing}.

The datasets in language models (PTB, WikiText) include UNK tokens which are replacements of low frequency based OOV (Out of Vocabulary) tokens.
This replacement often hurts domain-specific datasets like scientific articles or news, whose classification could rely mainly on OOV entities. Hence, for our dataset, instead of using UNK tokens, we use named entity-specific UNK tokens (instead of using UNK, we use PERSON-UNK, COUNTRY-UNK, PRODUCT-UNK and so on). 
Named entities are often a good indicator of trends and could contribute to succinct classification \cite{dewdney2012named} \cite{newman2006analyzing}.
For example, a news article pertaining to "lifestyle" could refer to products and an article belonging to international relations could refer to countries. 
As both 'product' and 'country' are entities, such entity-specific tokens could improve our results. 
The information regarding named entities is done by preprocessing through SpaCy \cite{spacy2}.

We tested this hypothesis on our dataset: 
a well-tuned CNN gives 62.4\%, regular ULMFiT gives 73.5\%, and named-entities based ULMFiT gives 77.4\% accuracies on test dataset.
We deployed our model using PyTorch \cite{paszke2017pytorch}.

It should be noted that deciding news category for the article is often ambiguous and fuzzy (for example, an article about Apple could be categorized as "business" or "technology"). As such, our intent in category labeling is not to achieve perfect accuracy, but to facilitate better keyphrase extraction, as described in the next section.

\section{Keyphrase Extraction}
\begin{figure}[t]
    \centering
    \includegraphics[width=0.8\textwidth]{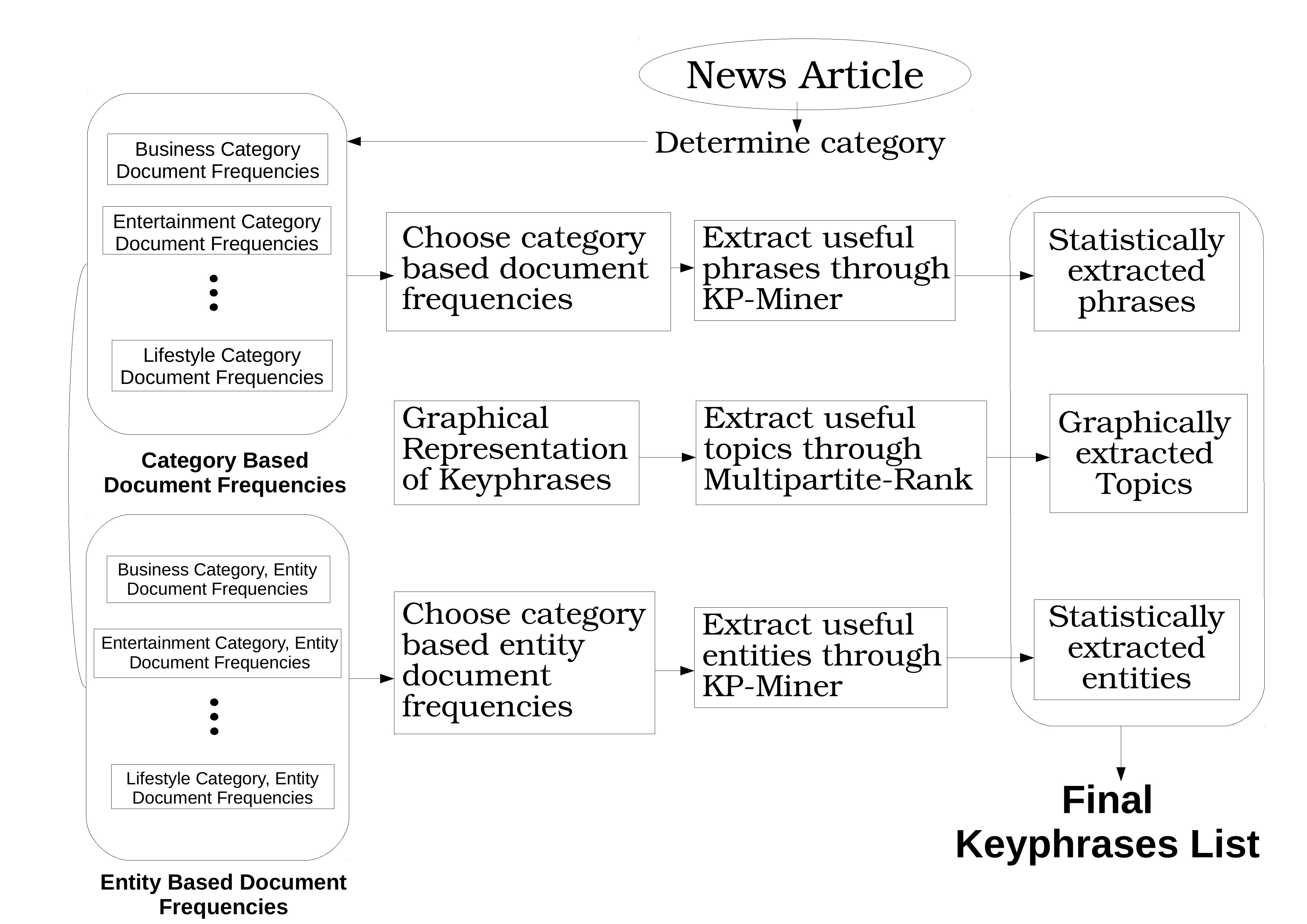}
    \caption{Schematic diagram for our keyphrase extraction module. Our keyphrase involves aggregation of statistically extracted phrases and entities, and graphically extracted topics.}
    \label{fig}
\end{figure}

The schematic diagram for keyphrase extraction module is shown in figure \ref{fig}.
We have used PKE (Python Keyphrase Extraction) library for keyword extraction \cite{boudin2016pke}.

It should be noted that vocabulary styles are different across news categories. 
Thus, it is crucial to have category-specific corpora.
Our extraction system relies on three different methods as follows:

\textbf{Statistically Extracted Keyphrases:}
The goal of this task is to extract keyphrases which are determined statistically.
KP-Miner serves this purpose by making use of document frequencies, term frequencies, and positional occurrences \cite{el2009kp}.
Once the article category is defined by the classifier, we use category-specific document frequencies for better retrieval.
We found that it was able to recommend phrases well; however, most of the single-word keywords were redundant.
Hence, we only use this module for multi-word keyphrases.

\textbf{Statistically Extracted Entities:}
The goal of this task is to extract the entities (like persons, organizations, places) which are central to the article.
To this end, we also use KP-Miner, but on entity-based category-specific document frequencies.
After extraction, we de-duplicate using singularization and "nounification".
Nounification of a keyword is performed by retrieving a relevant noun in the synset of the corresponding keyword in WordNet \cite{miller1995wordnet}.
This easily converts words like "elect" to "election" and "Italians" to "Italy".

\textbf{Graphically Extracted Topics:}
Graphical-based methods for keyword extraction comprises topics as nodes and edges as semantic relations. 
The aim of this task is to extract the key-topics which influence other surrounding topics in an article.
Many unsupervised approaches exist, like SingleRank \cite{wan2008single}, TopicRank \cite{bougouin2013topicrank}, TopicalPageRank \cite{sterckx2015topical}, PositionRank \cite{florescu2017positionrank} and MultipartiteRank \cite{boudin2018unsupervised}, among which MultipartiteRank gives the best result for this purpose.
However, MultipartiteRank often returns redundant keyphrases which are eliminated by de-duplication and nounification.

We aggregate the results from these three methods and display them as tags after de-duplication.
\section{Discussions and Further Steps}
The Inflo labeling system can be used to improve the efficiency of finding relevant content within an aggregation platform. Because classification is instantaneous and automatic, news articles shared from any source can be analyzed and labelled, making them easily accessible through the use of content navigation tools such as topic and sub-topic filters. Such filters could allow for a more personalized feed of relevant content, rather than a single stream of potentially irrelevant or uninteresting content. 

With the keyphrase extraction, news articles which are directly related (i.e. articles from a different source on a singular event / incident) could be clustered and presented together, making it possible to expand one’s perspective and take-in different viewpoints. 

Overall, the Inflo news labeling system, with its highly accurate output of insightful terms could expedite the process of finding relevant and interesting content on the web when implemented in an aggregation platform.

\newpage